# STRETCHED EXPONENTIAL DISTRIBUTIONS IN NATURE AND ECONOMY:

## "FAT TAILS" WITH CHARACTERISTIC SCALES

**Jean LAHERRERE** [1] and **Didier SORNETTE** [2]

[1]107 rue Louis Blériot, 92100 Boulogne,     e-mail : j.h.laherrere@infonie.fr

[2]Laboratoire de Physique de la Matière Condensée CNRS UMR 6622
and Université de Nice-Sophia Antipolis, B.P. 71, 06108 Nice Cedex 2, France
and Institute of Geophysics and Planetary Physics and Department of Earth and Space
Sciences, UCLA, Box 951567, Los Angeles, CA 90095-1567, USA,
        e-mail : sornette@cyclop.ess.ucla.edu

**Abstract**
To account quantitatively for many reported "natural" fat tail distributions in Nature and
Economy, we propose the stretched exponential family as a complement to the often
used power law distributions. It has many advantages, among which to be economical
with only two adjustable parameters with clear physical interpretation. Furthermore, it
derives from a simple and generic mechanism in terms of multiplicative processes. We
show that stretched exponentials describe very well the distributions of radio and light
emissions from galaxies, of US GOM OCS oilfield reserve sizes, of World, US and
French agglomeration sizes, of country population sizes, of daily Forex US-Mark and
Franc-Mark price variations, of Vostok temperature variations, of the Raup-Sepkoski's
kill curve and of citations of the most cited physicists in the world. We also briefly
discuss its potential for the distribution of earthquake sizes and fault displacements and
earth temperature variations over the last 400 000 years. We suggest physical
interpretations of the parameters and provide a short toolkit of the statistical properties
of the stretched exponentials. We also provide a comparison with other distributions,
such as the shifted linear fractal, the log-normal and the recently introduced parabolic
fractal distributions.

PACS: 02.50.+s : Probability theory, stochastic processes and statistics

Short title: Stretched exponential distributions

### 1-Introduction

Frequency or probability distribution functions (pdf) that decay as a power law of their argument

$$P(x) \, dx = P_0 \, x^{-(1+\mu)} \, dx \qquad (1)$$

have acquired a special status in the last decade. They are sometimes called ``fractal'' (even if this term is more appropriate for the description of self-similar geometrical objects rather than statistical distributions). A power law distribution characterizes the absence of a characteristic size : *independently* of the value of x, the number of realizations larger dans λx is $\lambda^{-\mu}$ times the number of realizations larger than x. In contrast, an exponential for instance or any other functional dependence does not enjoy this self-similarity, as the existence of a characteristic scale destroys this continuous scale invariance property [1]. In words, a power law pdf is such that there is the same proportion of smaller and larger events, whatever the size one is looking at within the power law range.

The asymptotic existence of power laws is a well-established fact in statistical physics and critical phenomena with exact solutions available for the 2D Ising model, for self-avoiding walks, for lattice animals, etc. [2], with an abondance of numerical evidence for instance for the distribution of percolation clusters at criticality [3] and for many other models in statistical physics. There is in addition the observation from numerical simulations that simple ``sandpile'' models of spatio-temporal dynamics with strong non-linear behavior [4] give power law distributions of avalanche sizes. Furthermore, precise experiments on critical phenomena confirm the asymptotic existence of power laws, for instance on superfluid helium at the lambda point and on binary mixtures [5]. These are the ``hard'' facts.

On the other hand, the relevance of power laws in Nature is less clear-cut even if it has repeatedly been claimed to describe many natural phenomena [1,6,7]. In addition, power laws have also been proposed to apply to a vast set of social and economic statistics [8-14]. Power laws are considered as one of the most striking signatures of complex self-organizing systems [15,16]. Empirically, a power law pdf (1) is represented by a linear dependence in a double logarithmic axis plot (a log-log plot for short) of the frequency or cumulative number as a function of size. However, logarithms are notorious for contracting data and the qualification of a power law is not as straightforward as often believed. Claims have thus been made on the power law dependence of many data that have or will probably not survive a closer scrutiny. See for instance [17,18] which point out that disorder and irregular boundaries may lead to apparent scaling over two decades. See also [19] which points out that the scaling range of experimentally declared fractality in laboratory experiments is extremely limited as most log-log plots show a straight portion over only 1.3 order of magnitude.

In general, in all power laws of critical phenomena, we observe them only asymptotically, i.e. with infinite purity, infinite temperature control, infinitely large computer simulations, waiting infinitely long for equilibration, and making gravity effects infinitely small by going into space. This was called Asymptopia by R.A. Ferrell about thirty years ago. In any real experiment or simulation, there are deviations from Asymptopia and thus deviation from power laws. Our paper is **not** about these unavoidable but reducable observation errors, but claims that even infinitely accurate experimental preparations and observations in our examples treated below give important deviations from power laws.

Thus, in any finite critical system, it is well-known that the power law description must give way to another regime dominated by finite size effects [20] and the pdf's in general cross over to an exponential decay, which leads to a curvature in the log-log plots. Log-log plots of data from natural phenomena in Nature and Economy often exhibit a limited linear regime followed by a significant curvature. The outstanding question is whether these observed deviations from a power law description result simply from a finite-size effect or does it invade the main body of the distribution, thus calling for a more fundamental understanding and also a completely different quantification of the pdf's. The first hypothesis has often been suggested for instance for the Gutenberg-Richter distribution of earthquake sizes: if the power law distribution was extrapolated to infinite sizes, it would predict an infinite mean rate of energy release, which is clearly ruled out in a finite earth. Similarly, the extrapolation of the distribution of the dispersed habitat of oilfield reserves gives an infinite quantity of oil. This is clearly ruled out in a finite earth and a cross-over to another regime is called for. The fact that most of the natural distributions display a log-log curved plot [21], avoiding the divergence and leading to thinner tails than predicted by a power law, has mostly been interpreted in terms of finite-size effects.

Here, we explore and test the hypothesis that the curvature observed in log-log plots of distributions of several data sets taken from natural and economic phenomena might result from a deeper departure from the power law paradigm and might call for an alternature description over the **whole range** of the distribution. The choice of a given mathematical class of distributions corresponds to a ``model'' in the following sense. In its broadest sense, a model is a mathematical representation of a condition, process, concept, etc, in which the variables are defined to represent inputs, outputs, and intrinsic states and inequalities are used to describe interactions of the variables and constraints on the problem. In theoretical physics, models take a narrower meaning, such as in the Ising, Potts,..., percolation models. For the description of natural and of economic phenomena, the term model is usually used in the broadest sense that we take here.

The model that we test is provided by the recent demonstration that the tail of pdf's of products of a finite number of random variables is generically a stretched exponential [22], in which the exponent c is the inverse of the number of generations (or products) in a multiplicative process. We thus propose an alternative model for pdf's for natural and economic distributions in terms of stretched exponentials :

$$P(x) \, dx = c \, (x^{c-1}/ \, x_0^c) \, \exp[-(x/x_0)^c] \, dx \, , \tag{2}$$

such that the cumulative distribution is

$$P_c(x) \; = \exp[-(x/x_0)^c] \; . \tag{3}$$

Stretched exponentials are characterized by an exponent c smaller than one. The borderline c=1 corresponds to the usual exponential distribution. For c smaller than one, the distribution (3) presents a clear curvature in a log-log plot while exhibiting a relatively large apparent linear behavior, all the more so, the smaller c is. It can thus be used to account both for a limited scaling regime and a cross-over to non-scaling. When using the stretched exponential pdf, the rational is that the deviations from a power law description is fundamental and not only a finite-size correction.

We find that the stretched exponential (2-3) provides an economical description as it depends on only two meaningful adjustable parameters with clear physical interpretation (the third one being an unimportant normalizating factor). It accounts very well for the distribution of radio and light emissions from galaxies (figure 2), of US GOM OCS oilfield reserve sizes (figures 3 & 4), of World, US and French agglomeration sizes (figures 5-8), of the United Nation 1996 country sizes (figure 9), of daily Forex US-Mark and Franc-Mark price variations (figures 10 & 11), and of the Raup-Sepkoski's kill curve (figures 12), Even the distribution of biological extinction events [23] is much better accounted for by a stretched exponential than by a power law (linear fractal). We also show that the distribution of the largest 1300 earthquakes in the world from 1977 to 1992 (figures 13) and the distribution of fault displacements (figures 14) can be well-described by a stretched exponential. We also study the tempeture variations over the last 420 000 years obtained for ice core isotope measurements (figures 15). Finally, we examine the distribution of citations of the most cited physicists in the world and again find a very fit by a stretched exponential (figures 16).

We do not claim that all power law distributions have to be replaced but that observable curvatures in log-log plots that are often present may signal that another statistical representation, such as a stretched exponential, is better suited. This in turn may inspire the identification of a relevent physical mechanism. Among the possible candidates, stretched exponentials are particularly appealing because, not only do they provide a simple and economical description but, there is a generic mechanism in terms of multiplicative processes. Multiplicative processes often constitute zeroth-order descriptions of a large variety of physical systems, exhibing anomalous pdf's and relaxation behaviors.

Stretched exponential laws are familiar in the context of anomalous relaxations in glasses [24] and can be derived from a multiplicative process [22]. The Ising ferromagnet in two dimensions was also predicted to relax by a stretched-exponential law [25]. This was confirmed numerically [26] and by improved arguments [27]. It has been claimed that the effect also exists in three dimensions [28] but is not confirmed numerically (D. Stauffer, private communication). The theory is based on the existence of very rare large droplets (heterophase fluctuations). Ising models deal with interacting spins and thus dependent random variables. This may be transformed into independent variables by considering suitable groups of spins (the droplets), as can be done for variables with long range correlation [29], and then the theory of the extreme deviations for the product of independent random variables can apply [22]. This might suggest a deep link between the anomalous relaxation in the Ising model and the data sets that we analyze here. Let us also mention that, up to the space dimension 3+1 for which numerical simulations have been carried out [30], the distribution of heights in the Kardar-Parisi-Zhang equation of non-linear stochastic interface growth is also found to be a stretched exponential.

We present the different data sets in the next sections and provide in the appendix a short toolkit calculus for the statistical analysis of stretched exponentials. The qualification of stretched exponentials is particularly important with regards to extrapolations to large events that have not yet been observed.

## 2-Evidence of stretched exponentials and comparison with other laws

In our analysis, we use the rank-ordering technique [9,31,32] that amounts to order the variables by descending values $Y_1 > Y_2 > ... > Y_N$, and plot $Y_n$ as a function of the rank n. Rank-ordering statistics and cumulative plots are equivalent except that the former provides a perspective on the rare, largest elements of a population, whereas the statistics of cumulative distributions are dominated by the more numerous small events. As a consequence, statistical fluctuations describe the uncertainties in the values of $Y_n$ for a given rank in the rank-ordering method while they describe the variations of the number of events of a given size in the cumulative representation. This shows that the former statistics is better suited for the analysis of the tails of pdf's characterized by relatively few events.

Within the rank-ordering plot, a stretched exponential is qualified by a straight line when plotting $Y_n^c$ as a function of log n, as seen from eq.(3). A fit of the straight line gives

$$Y_n^c = - a \ln n + b \qquad (4)$$

corresponding to

$$x_0 = a^{1/c} . \qquad (5)$$

Indeed, the definition (3) of the cumulative distribution $P_c(x) = \exp[-(x/x_0)^c]$ for the stretched exponential means that $\ln P_c(x) = -(x/x_0)^c$. Since $P_c(x) = n/N$, where n is the number of events larger or equal to x, hence n is the rank, we obtain (4) immediately.

The analysis of [22] predicts that multiplicative processes lead to a stretched exponential of the form

$$P_c(x) = \exp[- m (x/\sigma)^{1/m}/\lambda] \ , \tag{6}$$

where the index 'c' stands for the cumulative distribution, defined by expression (3). Here,

$$m=1/c \tag{7}$$

is the number of levels in the multiplicative cascade. $\sigma$ is the unit of the variable x, such that $x/\sigma$ is dimensionless. $\lambda$ is a typical multiplicative factor defined by the fact that $x/\sigma$ is the product of m random dimensionless variables of typical size $\lambda$. We see that a fit to a data set by a stretched exponential gives access only to the product $\sigma^{1/m} \lambda$ by the relation $x_0 = c\, \sigma^c\, \lambda$.

In order to interpret correctly the meaning of $x_0$, the appendix shows that the mean $<x>$ is given by

$$<x> = x_0 \ \Gamma(1/c) / c \ , \tag{8}$$

where $\Gamma(x)$ is the gamma function (equal to (x-1)!  for x integer). When c is small, $<x>$ will be much larger than $x_0$. $x_0$ is thus not the typical scale of x, but a reference scale from which all moments can be determined. Another characteristic scale $x_{95\%}$ can be obtained such that the probability to exceed this value is less than 5% (corresponding to a 95% level confidence). Using (6), this yields

$$x_{95\%} = 3^{1/c} \, x_0 \ . \tag{9}$$

Notice that the figures and the fits are done using the decimal logarithm. We thus convert the values of the fits to the natural logarithm to estimate these numbers.

The intuitive interpretation of the three parameters a, b and c of the stretched exponential rank ordering fit (4) is the following: $b^{1/c}$ is the size of the event of rank 1 (i.e. the largest event in the population), $x_0 = a^{1/c}$ according to eq.(5) is a characteristic scale from which one can deduce the value of the mean and of various moments as seen from eqs.(8) and (9). Finally, the exponent c quantifies the fatness of the tail of the stretched exponential, the smaller the exponent, the fatter the tail. Within the multiplicative model [22], its inverse is proportional to the number of generations.

**Figure 1** compares the stretched exponential model with three other models also exhibiting a curvature in the fractal display (log-log size-rank). These three models are the following. The first one corresponds to a simple parabola in the log-log plot, and is called the parabolic fractal [21],

$$\log S_n = \log S_1 - a \log n - b \, (\log n)^2 \ . \tag{10}$$

In expression (10), $S_n$ is the size of rank n. The case b=0 recovers the usual linear log-log plot qualifying a pure power law distribution. The introduction of the quadratic term $-b \,(\log n)^2$ is a natural parametric addition to account for the existence of curvature in the log-log plot. The three parameters $\log S_1$, a and b of the fit with eq.(10) have simple interpretations: $\log S_1$ is the (decimal) logarithm the size of the event of rank 1 (i.e. the largest event in the population), a is the slope (inverse of the power law exponent $\mu$) for the largest values (smallest ranks) and b quantifies the curvature while $1/(a+b \log n)$ is the apparent power law exponent. Solving this quadratic equation for n as a function of $S_n$, we find the corresponding cumulative distribution function (cdf) giving the number of events n larger than $S_n$:

$$n = n_0 \exp[(1/\sqrt{b}) \{\log(S_{max}/S_n)\}^{1/2}] , \qquad (11)$$

where

$$S_{max} = S_1 \exp[a^2/4b] , \qquad (12)$$

and

$$n_0 = e^{-a/2b} . \qquad (13)$$

The most remarkable property of this parabolic fractal distribution is the existence of a maximum $S_{max}$ beyond which the distribution does not exist. In other words, the pdf and cdf of the parabolic fractal have finite compact support. This is different from a finite-size effect leading to a maximum size in a finite system as here the maximum size remains finite even in the infinite asymptotic case. This property contrasts the parabolic fractal from the other distributions that are defined for arbitrarily large arguments.

The second model is the curved shifted linear fractal

$$\log S_n = \log S_0 - a \log (n+A) , \qquad (14)$$

and the last model we consider is the lognormal distribution (with standard deviation s and most probable value m). We have chosen the parameters of these four models in such a way that they approach each other the most closely in the interval of ranks 5-500, as shown in **Figure 1.**

All these models need three parameters, one for normalization and two that characterize the shape. As for the weights of small events in this numerical simulation which takes as element of comparison the overlap from rank 5 to 500, the smallest is the lognormal, then the stretched exponential, then the parabolic fractal and last the shifted linear fractal (which is completely linear beyond rank 100) in this numerical example. For the tail towards large events, the thinner distribution is the parabolic fractal (since it is limited), then the stretched exponential, then the lognormal and last the shifted linear fractal.

### 2-1 Radio and light emissions from galaxies

There is an undergoing controversy on the nature of the spatial distribution of galaxies and galaxy clusters in the universe with recent suggestions that scale invariance could apply [33-35]. Motivated by this question and the relationship between galaxy luminosities and their spatial patterns [36], we investigate the pdf of radio and light intensities radiated by galaxies. The data is obtained from [37]. The problem is that it is difficult to obtain a complete distribution of the global universe as the observation are done by sectors and the universe is not homogeneous. It has been shown [21] that the rank-ordering plot in log-log coordinates is not strictly linear for these data sets and that a noticeable curvature exists. **Figure 2** shows the radio intensity of the galaxies raised to the power c=0.11 as a function of the (decimal) logarithm of the rank n and the light intensity of the galaxies raised to the power c=0.04 as a function of the (decimal) logarithm of the rank n. The curves are convincingly linear over more than five decades in n.

The best fit to the data for radiosources gives $x_0 \sim 4 \; 10^{-8}$ for the radio emissions. From expression (8), $\langle x \rangle \approx 9! \; x_0 \approx 1.6\text{-}3.2 \; 10^{-2}$. Expression (9) gives $x_{95\%} \sim 9 \; 10^{-4}$. The best fit to the data gives $x_0 \sim 2 \; 10^{-34}$ for the light emissions. . From expression (8), $\langle x \rangle \approx 25! \; x_0 \approx 1.6 \; 10^{25} x_0 \approx 3 \; 10^{-9}$. Expression (9) gives $x_{95\%} \sim 2 \; 10^{-22}$. In these cases where the exponent c is very small, the reference scale $x_0$ is very small compared to a typical scale obtained from $\langle x \rangle$ or $x_{95\%}$. The large uncertainties are due to the smallness of the exponent c whose inverse is thus poorly constrained.

Theories of these radio and light emissions are poorly constrained but, interpreted within the multiplicative cascade model [22], this indicates of the order of 10-20 (resp. 25-50) levels in the cascade for radio (resp. light) emissions. The factor two in the range of this estimation of the number of levels stems from the choice of an exponential versus a Gaussian for the distribution of the variables that enter in the multiplicative process (see [22] for more details).

### 2-2 US GOM OCS oil reserve sizes

The determination of the distribution of oil reserves in the world is of great significance for the assessment of the sustainability of energy consumption by mankind (and especially by developed countries) in the future. The extrapolation of the statistical data may bring useful inference on the rate of new discoveries that can be expected in the future. It is thus particularly important to correctly characterize the distribution. The data are usually confidential and politically sensitive in OPEC countries where quotas depend of reserves. Most of reserve databases are unreliable. We have chosen the public data of the US OCS (outer continental shelf) published in open file by MMS (Mineral Mangement Service of the US Department of Interior). We have found that it is important to select the data within one unique natural domain. For oil reserves, the domain is the Petroleum System defined by its source-rock, i.e the genetic origin when most of previous studies were carried out on tectonic classification.

In ref.[21], it was noticed that the log-log rank-ordering plot exhibits a sizable downward curvature, indicating a significant deviation from a power law distribution. **Figure 3** shows the rank-ordering plot of oil field sizes (in million barrels = Mb) in a log-log display by decades. It is obvious that the largest fields were found first and that the leftmost part of the curved plot for the smallest ranks did not change during the last twenty years. This part of the curve can be easily extrapolated with a parabola which aim to represent the ultimate distribution of oil reserves in the ground encompassing all smaller oil fields, many of those that have probably not yet been discovered.

**Figure 4** represents the same data raised to the power c=0.35 as a function of the (decimal) logarithm of the rank n. An excellent fit to a straight line is found over almost three decades in ranks which provides a characteristic size $x_0 = 3 \pm 1$ using eq.(5) and $<x> \approx x_{95\%} \approx 80$ using expressions (8) and (9).

Interpreted within the cascade multiplicative model, the value of the exponent $c \approx 1/3$ corresponds to about 3 to 6 generation levels in the generation of a typical oil field. On the same display, we have plotted the ultimate curve of the figure 3 and obtain a linear fit for c=0.21. The ultimate curve is obtained by fitting the parabolic fractal law to the first ranks in fig.3. In this way, we get a plausible asymptotic corresponding to the reliable knowledge of all oil reservoirs in the earth. In this case where the largest sizes are well known and the small ones hidden, it is easier to extrapolate the plot in a log-log display than in a log-power. The stretched exponential may thus give us a clue as to the formation of the oil fields but the parabolic fractal is probably better suited for the extrapolation towards small oil fields. In particular, the existence of a maximum size that characterizes the parabolic fractal distribution is well-adapted to account for this real data set.

### 2-3 World, US and French urban agglomeration size distributions and UN population per country

Urban agglomerations provide an example of self-organization that has recently been studied from the point of view of statistical physics of complex growth patterns [38]. A model has been proposed [39] in terms of strongly correlated percolation in a gradient that accounts reasonably well for the morphology of large cities and their growth as a function of time. The distribution of areas occupied by secondary towns around a major metropole has been found approximately to be a power law [39]. Here, we analyze the distribution of urban agglomeration sizes. The term agglomeration refers to the natural geographic limits (defined by the continuity of the buidings (no gap larger than 200 m) as opposed to the artificial administrative boundaries that give spurious results. We take the population as the proxy for the agglomeration size.

**Figure 5** shows the rank-ordering plot of agglomeration sizes (larger than 100 000 inhabitants) in the US (United Nation demographic yearbook 1989) raised to the power c=0.165 as a function of the (decimal) logarithm of the rank n. An excellent fit to a straight line is found over more than two decades in ranks which provides a reference scale $x_0 = 20 \pm 2$ using eq.(5) and the typical agglomeration size $<x> \approx x_{95\%} \approx 15$ 000, using expressions (8) and (9).

**Figure 6** displays the data on a log-log format with the best fit with the different laws discussed above. The shifted linear freactal model is taken from the book of Gell-Man [15]. We use the cumulative distributions for each of the models to get the rank ordering plot. The best fit for each model is extrapolated from the city size of 100 000 inhabitants down to the extreme minimum size of one person to test which model predict the best the total size of the population in cities of size less than 100 000 inhabitants. The total population in 1989 of the US was 243 million. The 258 agglomerations over 100 000 people account for 187 million persons. The rest, which represents 56 million persons, live in cities smaller than 100 000 inhabitants. The extrapolations of the different models provide a prediction for this number of people in cities smaller than 100 000 inhabitants. The stretched exponential predicts 28 million, the parabolic fractal predicts 46 million and the shifted linear fractal predicts 97 million. In this case, the best fit is obtained with the parabolic fractal distribution. It is not surprising that the stretched exponential underperforms as it is justified theoretically from [22] strictly for large events. We are nevertheless amazed by how the stretched exponential can usually account for the region of pdf's far from the extreme, even in the center of the distribution.

**Figure 7** shows the rank-ordering plot of agglomeration sizes (larger than 100 000 inhabitants) in France raised to the power c=0.18 (notice the robustness of the exponent compared to the US case) as a function of the (decimal) logarithm of the rank n. An excellent fit to a straight line is found over about two decades in ranks which provides a reference scale $x_0 = 7 \pm 1$ using eq.(5) and the typical agglomeration size $<x> \approx 288$ $x_0 = 1800$ and $x_{95\%} \approx 2900$, using expressions (8) and (9). Notice the existence of an outlier, Paris, which is much above the extrapolation of the straight line. This has been coined the ``king'' effect (the new King kills the barons to avoid competition and to acquire a wealth above the commoners!) [21].

**Figure 8** shows the rank-ordering plot of agglomeration sizes (larger than 100 000 inhabitants) in the world [40] raised to the power c=0.13 (notice again the relative robustness of the exponent compared to the US and French case) as a function of the (decimal) logarithm of the rank n. An excellent fit to a straight line is found over more than four decades in ranks notwithstanding the fact that the database is probably not perfect as several hundreds of people are missing from the counting in crowded countries like China and India. The fit shown in the figure provides a reference scale $x_0 = 0.03$ using eq.(10) and the typical agglomeration size $<x> \approx 2.1 \ 10^4$ $x_0 = 600$ and $x_{95\%} \approx 120$, using expressions (8) and (9). The fact that $<x>$ is much larger than $x_0$ reflects the fat tail nature of the distribution with the small exponent c.

**Figure 9** shows the rank-ordering plot of country population sizes in the world reporded by the United Nations (Urban agglomeration 1996). Each country population has been raised to the power c=0.42. A good fit to a straight line is found over more than two decades in ranks. The fit shown in the figure provides a reference scale $x_0 = 7$ million using eq.(10) and the typical country size $<x> \approx 19.5$ million and $x_{95\%} \approx 91$ million, using expressions (8) and (9). Notice the existence of two outliers or ``kings'', China and India.

### 2-4 Daily Forex US-Mark and US-Franc price variations

Stock market prices fluctuate under the action of many factors and the precise characterization of the distribution of price variations has important applications for option pricing, portfolio optimization and trading. In addition, from a theoretical point of view, it constraints the models of the stock market. Historically, the central limit theorem led to the first paradigm in terms of Gaussian pdf's that was first put in doubt by Mandelbrot [41] when he proposed to use Lévy distributions, that are characterised by a fat tail decaying as a power law with index $\mu$ between 0 and 2. Recently, physicists have characterised more precisely the distribution of market price variations [42,43] and found that a power law truncated by an exponential provides a reasonable fit at short time scales (much less than one day), while at larger time scales the distributions cross over progressively to the Gaussian distribution which becomes approximately correct for monthly and larger scale price variations. Alternative representations exist in terms of a superposition of Gaussian pdf's corresponding to cascade models inspired from an analogy with turbulence [44]. These two classes of descriptions can only be distinguished using higher order statistics that seem to favor the cascade description [45].

The daily time scale is the most used for practical applications but is unfortunately fully in the cross-over regime between the truncated Lévy law at the shortest time scales and the asymptotic Gaussian behavior at the largest time scales. It has thus been poorly constrained. Here, we show that a stretched exponential pdf provides a parsimonious and accurate fit to the full range of currency price variations at this daily intermediate time scale. We present two different data dealing with foreign exchange rates. The foreign exchange market is slightly different from the other financial markets such as the stock market for instance since one does not exchange a valuable against money, but a currency for another currency. The foreign exchange is the most active market in the world with a daily turnover of more than a trillion US dollars, with a large part of the trades being done for hedging and speculative purposes. These transactions concern only some major currencies with the US dollar implied in 80% of them. The deutschmark (DEM) is the second major currency with 20% of the transactions are exchanges between US dollars and DEM. Contrary to the other financial markets, the foreign exchange market is a 24 hours global market. See [46] for more informations.

The first data is represented in **figure 10** which shows the positive and negative variations of the US dollar expressed in German marks during the period from 1989 to the end of 1994. We use again the rank-ordering representation and plot in **figure 10a** the nth price variation (positive with square symbols and negative with diamond symbols) in log-log coordinates. **Figure 10b** plots the nth price variation (positive with square symbols and negative with diamond symbols) taken to the power c=0.87 and 0.90 respectively as a function of the decimal logarithm of the rank. We observe an excellent description by the almost same straight line over the full range of quotation variations for both the positive and negative variations. This shows that the pdf is approximately symmetric: there is essentially the same probability for an appreciation or a depreciation of the US dollar with respect to the German mark. Notice that the apparent slight deviations above the straight line for the largest variations are completely within the expected error bars. The best fit to both positive and negative variations with eq.(4) gives the same consistent values a=0.008 and b=0.06 $\pm$ 0.005. From this, we obtain $x_0 = 0.5$ %, and from expressions (8) and (9), we get $<x> \approx 1.09\ x_0 \approx 1$ % and $x_{95\%} \approx 1.7$ %.

**Figure 11a** plots the nth price variation for the French Franc expressed in German marks (in the period from 1989 to the end of 1994) with the positive variations represented with square symbols and the negative variations represented with diamond symbols) in log-log coordinates. **Figure 11b** plots the nth price variation (positive with square symbols and negative with diamond symbols) taken to the power c=0.72 and 0.64 respectively as a function of the decimal logarithm of the rank. We observe an excellent description with straight lines over the full range of quotation variations. Two important differences with the $/Mark case are noteworthy. First, the exponent c is smaller, which corresponds to a ``fatter'' tail, i.e. the existence of larger variations. The dynamics of Franc/Mark exchange rate is thus wilder than that of the two stronger currencies $/Mark. Secondly, the coefficients a and b of the positive and negative exchange rate variations are different, characterizing a clear asymmetry with larger negative variations of the Franc expressed in Marks. This asymmetry corresponds to a progressive depreciation of the Franc with respect to the Mark. One could have imagined that such a depreciation would correspond to a steady drift on which are superimposed symmetric variations. We find something else: the depreciation is putting its imprints at all scales of price variations and is simply quantified, not by a drift, but by a different reference scale $x_0$.

The best fit to the positive variations of the Franc/Mark exchange rate with eq.(4) gives a=0.014 and b=0.10. From this, we obtain $x_0 = 0.12$ %, and from expressions (8) and (9), we get $<x> \approx 1.4 \ x_0 \approx 0.17$ % and $x_{95\%} \approx 5.6 \ x_0 \approx 0.7\%$. The difference between $<x>$ and $x_{95\%}$ illustrates clearly the wilder character of the fat tail of the Frank-Mark exchange rate variations compared to the \$/Mark.

The best fit to the negative variations of the Franc/Mark exchange rate with eq.(4) gives a=0.0095 and b=0.07. From this, we obtain $x_0 = 0.16$ %, and from expressions (8) and (9), we get $<x> \approx 0.2$ % and $x_{95\%} \approx 4.6 \ x_0 \approx 0.7\%$. The difference between $<x>$ and $x_{95\%}$ illustrates clearly the wilder character of the fat tail of the Frank-Mark exchange rate variations compared to the \$/Mark and the fact that the depreciation of the Franc can occur by large and sudden drops rather than according to a steady drift.

To sum up this subsection, we have found that the stretched exponential quantifies in a remarkably simple and illuminating way the difference between the exchange rate between two strong currencies and between a strong and a weaker currency. This quantification uses only the two adjustable parameters c and $x_0$ that represent the fatness of the tail of the stretched exponential pdf and its reference scale.

### 2-5 Raup-Sepkoski's kill curve

It has been argued that the histogram of biological extinction events over the last 600 million years obtained from the fossil record ``can be reasonably well fitted to a power law with exponent between 1 and 3'' (see [16] p.165). **Figure 12a** reproduces the data from Sepkoski's compilation [47] in the log-log plot of the cumulative distribution with inverted axis (a rank-ordering plot). Notice that the rank does not start at n=1 but at the rank of the order of 60 because there are about 60 Genera that have a life span larger than 150 millions years and the data are not precise enough to distinguish between these 60 Genera. The plot does not show any straight line whatsoever but rather a systematic downward curvature (very close to a parabola) implying a much ``thinner'' tail than predicted by a power law. The fit is carried out using the parabolic fractal given by eq.(10). The fit is very good and suggest that there might be a maximum lifespan of about 350 millions years for species as predicted from eq.(12). **Figure 12b** shows the alternative rank-ordering plot of the lifespans measured in million years raised to the power c=0.85 as a function of the (decimal) logarithm of the rank n.

The curve is less convincing than for the previous examples as it exhibits a sigmoidal shape, but the data are much more difficult to obtain and probably much less reliable. However, a straight line provides a reasonable fit over about two decades in ranks. This provides a reference scale $x_0 = 22$ million years using eq.(5) and the typical lifespans $<x> \approx 1.09 \ x_0 \approx 25$ million years and $x_{95\%} \approx 3.6 \ x_0 \approx 82$ million years, using expressions (8) and (9). These numbers seem very reasonable when looking directly at figure 39 of ref.[ 16] p.165.

### 2-6 Earthquake size and fault displacement distributions

The well-known Gutenberg-Richter law gives the number of earthquakes in a given region (possibly the world) with magnitude larger than a given value. Translated into seismic moment (roughly proportional to energy released), the Gutenberg-Richter law corresponds to a power law distribution (1) with an exponent $\mu$ close to 2/3. $\mu$ being smaller than 1, the average energy released by earthquakes is mathematically infinite, or in other words it is controlled by the largest events. There must thus be a cross over to another regime falling faster and several models have previously been discussed, in terms of another power law for the largest earthquakes with exponent μ larger than 1 [31,48] or in terms of a Gamma distribution corresponding approximately to an exponential tail [49]. It is thus sometimes found that the Gutenberg-Richter law is too much linear [50]! Here, we reexamine the worldwide Harvard catalog (see [31] for a description and an analysis in terms of rank-ordering with power law distributions) containing the 1300 largest earthquakes in the world from 1977 to 1992. **Figure 13a** shows the rank-ordering plot of the nth seismic moment as a function of the rank n in the usual log-log representation. A clear bending is observed that can be well-fitted by the parabolic fractal distribution (formula (10)). **Figure 13b** shows the rank-ordering plot of the nth seismic moment raised to the power c=0.1 as a function of the logarithm of the rank n. A fit by a stretched exponential distribution is also very good. Both models fit very well and are similar for the extrapolation towards the smallest events. The choice of one model is however of significant consequence for the prediction of the size of the next largest earthquake. The parabolic fractal is in the present case ill-suited as it predicts a maximum size close to the largest observed event. This is probably underestimated and a stretched exponential extrapolation is probably to be prefered.

The distribution of fault displacements has been studied quantitatively to go beyond the usual geometrical description and quantify the relative activity of faults within complex fault networks. Fault displacements provide a long-term measurement of seismicity and are thus important for seismic hazard assessment and long-term prediction of earthquakes [51]. In contrast to a widespread belief, the distribution is not a power law as seen in **figure 14a** which represents the rank-ordering plot of the nth largest displacement of seismic faults [52] as a function of the nth rank in log-log coordinates. The curvature is very strong and is fitted to the parabolic fractal. **Figure 14b** qualifies essentially an exponential distribution since the nth largest displacement raised to a power close to 1 is linear in the logarithm (decimal) of the nth rank. The characteristic displacement is essentially given by the coefficient a of the fit equal to about 400/ln(10) ≈ 180 meters.

### 2-7 Temperature data over 400 000 years from Vostok near the south pole

Isotope concentrations in ice cores measured at Vostok near the south pole provide proxies for the earth temperature time series over the last 400 000 years, with more than 2600 data points [53]. A large research effort is focused at improving the reliability of this proxy and analyzing it to detect trends and oscillatory components that might be useful for climate modelling and for the assessment of present temperature warming trends [54]. To prepare the following plots, we normalize the temperature variations by the corresponding time interval. **Figure 15a** represents the log-log rank-ordering plot of the nth largest normalized temperature variations (positive and negative variations are treated separately) as a function of the nth rank. The curvature is very strong and clearly excludes a power law distribution. **Figure 15b** shows that a stretched exponential distribution can account reasonably well for both the distribution of positive and negative variations. The difference in the exponent c for the positive and negative temperature variations is not statistically significant : c ≈ 0.65. The same holds true for the characteristic value $x_0$≈ 13. We obtain, from expressions (8) and (9) <x> = 17 and $x_{95\%}$ =70. From this one-point statistical analysis, there is not much difference between the positive and negative temperature variations, with however a perceptible tendancy for observing more often larger negative temperature variations (the curve for the negative variations is systematically above that for the negative variations). Over this 400 000 year period and within this one-point statistical analysis, the temperature trend, if any, is more towards cooling than warming. Higher-order statistics, taking into account correlations between successive times such as by studying the time series directly, are needed to ascertain any recent warming trend.

2.8 Citations of the 1120 most cited physicists over the period1981-June 1997

D. A. Pendlebury from the Institute for Scientific Information has recently ranked by total citations the 1120 most cited physicists over the period1981-June 1997. The data represent citations recorded over 1981-June 1997 to ISI-indexed physics papers 1981-June 1997, and do not represent citations to books, to pre-1980 papers indexed by ISI, or to any papers not indexed by ISI during 1981-June 97. **Figure 16a** shows in log-log plot the dependence of $S_n$ as a function of rank n, where $S_n$ is the total number of citations of the nth most cited physicist. One observes clearly a curvature both for the smallest and the largest ranks, excluding a power law distribution. If however one insists in fitting this curve with a power law, one gets an apparent average exponent $\mu$ around $1/0.36 \approx 3$. **Figure 16b** shows $S_n$ raised to the power c=0.3 as a function of the natural logarithm of the rank. The whole range (over three decades in rank) is well-described by the stretched exponential. From the parameters of the fit shown on the figure, we obtain $x_0 = 2.7$ using eq.(5) and the typical citation numbers $<x> \approx 25$ and $x_{95\%} \approx 105$. For the physicists (to whom the present authors belong) who do not appear in this aristocracy of the 1120 most cited, it should come as a small satisfaction that it suffices to have more than about 105 citations to be among the 5% most cited physicists in the world! The 1120th rank has 2328 citations (while the first rank has ten times more): reporting this number 2328 in eq.(3) and using the value $x_p = 2.7$, this shows that the 1120 most cited physicists correspond to the fraction $5 \ 10^{-4}$ of the total physicist population, a number that appears quite reasonable.

We propose to rationalize the stretched exponential again using the results of [22] for multiplicative processes. It appears reasonable to involve a multiplication of factors to account for the impact of a scientist. An author has a large citation impact if he/her has (1) the ability to select an good problem for investigation, (2) the competence to work on it and carry out the work to completion, (3) the ability to create or belong to a research group and make it work efficiently, (4) the ability of recognizing a surprising and worthwhile result, (5) gifts for writing clear and lively papers, (6) the expertise, salemanship and dedication to advertise his/her results. Similar ideas were put forward by Shochley [55] who analyzed in 1957 the scientific output of 88 research staff members of the Brookhaven National Laboratory in the USA. He found a log-normal distribution which is the center of the limit distribution for the product of a large number of random variables. The fact that the extreme tail of the distribution that we analyze here is a stretched exponential is of no surprise within this model in view of the extreme deviation theorem [22].

## 3- Conclusion

Power laws are generally used to represent natural distributions, often claimed to be power laws which represent as linear regressions in log-log plots. In reality however, the plots often display linearity over a limited range of scales and/or exhibit noticeable curvature. In some cases as in oil field reserves and earthquake sizes, the small value of the exponent would imply a diverging average, a result ruled out by the finite size of the earth. Here, we have investigated the relevance for a set of ten different data sets of the family of stretched exponential distributions. We have also compared the fits with a natural extension of the linear fits in log-log plots using a quadratic correction, which leads to the so-called parabolic fractal distribution. The stretched exponential seems to provide a reasonable fit to all the data sets and has the advantage of a sound theoretical foundation. We have been however surprised to realize that the stretched exponential pdf's, that are supposed theoretically to apply better for the rarest events, seem to account remarkably well for the center of most of the analyzed distributions. Stretched exponential have a tail that is ``fatter'' than the exponential but much less so that a pure power law distribution. They thus provide a kind of compromise between the two descriptions. Stretched exponentials have also the advantage of being economical in their number of ajustable parameters. The parabolic fractal is also a natural parametric representation, that sometimes perform better for the natural (non-economic) data sets and exhibits robustness in its parameters a and b.

**Acknowledgments**: We thank D. Stauffer for very useful comments and suggestions and D. A. Pendlebury from the Institute for Scientific Information for the data on physicist citations.

# Appendix

Using the parametrization (2)

$$P(x) \, dx = c \, (x^{c-1}/x_0^c) \, \exp[-(x/x_0)^c] \, dx \, , \qquad (A1)$$

we have

$$\langle x \rangle = x_0 \, (1/c) \, \Gamma(1/c) \, , \qquad (A2)$$

and

$$\langle x^2 \rangle = x_0^2 \, (2/c) \, \Gamma(2/c) \, , \qquad (A3)$$

where $\Gamma(x)$ is the Gamma function.

Let us now give the most probable determination of the parameters $x_0$ and $c$. Using the maximum likelihood method, we find

$$x_0^c = (1/n) \, \Sigma_{i=1}^{n} \, (Y_i^c - Y_n^c) \qquad (A4)$$

where $Y_1 > Y_2 > ... > Y_i > ... > Y_n$ are the n largest observed values. This expresssion (A4) provides the most probable value for $x_0$ conditioned on the knowledge of the exponent c. The method of maximum likelihood also allows us to get an equation for the most probable value of the exponent c :

$$1/c = [\Sigma_{i=1}^{n} \, (Y_i^c \, \ln Y_i - Y_n^c \, \ln Y_n)] / [\Sigma_{i=1}^{n} \, (Y_i^c - Y_n^c)] - (1/n) \, \Sigma_{i=1}^{n} \ln Y_i$$

which is implicit as c appears on both sides of the equality.

We now provide the distribution of extreme variations. We ask what is the probability $P(x_{max} > x^*)$ that the largest value $x_{max}$ among N realizations be greater than $x^*$ :

$$P(x_{max} > x^*) = 1 - [1 - \exp[-(x^*/x_0)^c]]^N \cong 1 - \exp\{-N \exp[-(x^*/x_0)^c]\} \, . \qquad (A5)$$

Denote p the chosen level of probability of tolerance for the largest value $x_{max}$, i.e. $P(x_{max} > x^*) = p$. Inverting (A5) yields

$$x^* = x_0 \, \{Log \, [-N/Log(1-p)]\}^{1/c} \, . \qquad (A6)$$

The following table is calculated for c=0.7, x0=0.027 and N=7500. The value x*= 0.062 is the usual estimate of the typical largest value, corresponding to a probability of 37%.

| p | 1-1/e=0.63 | 1/2 | 0.1 | 0.01 | 0.001 |
|---|---|---|---|---|---|
| $x^*/x_0$ | 22.9 | 24.2 | 31.8 | 41.5 | 52.2 |
| $x^*$ | 0.062 | 0.065 | 0.086 | 0.112 | 0.141 |

.

**figure 1 :**

## Comparison between parabolic fractal, shifted linear fractal, lognormal and stretched exponential fitted up to rank 500

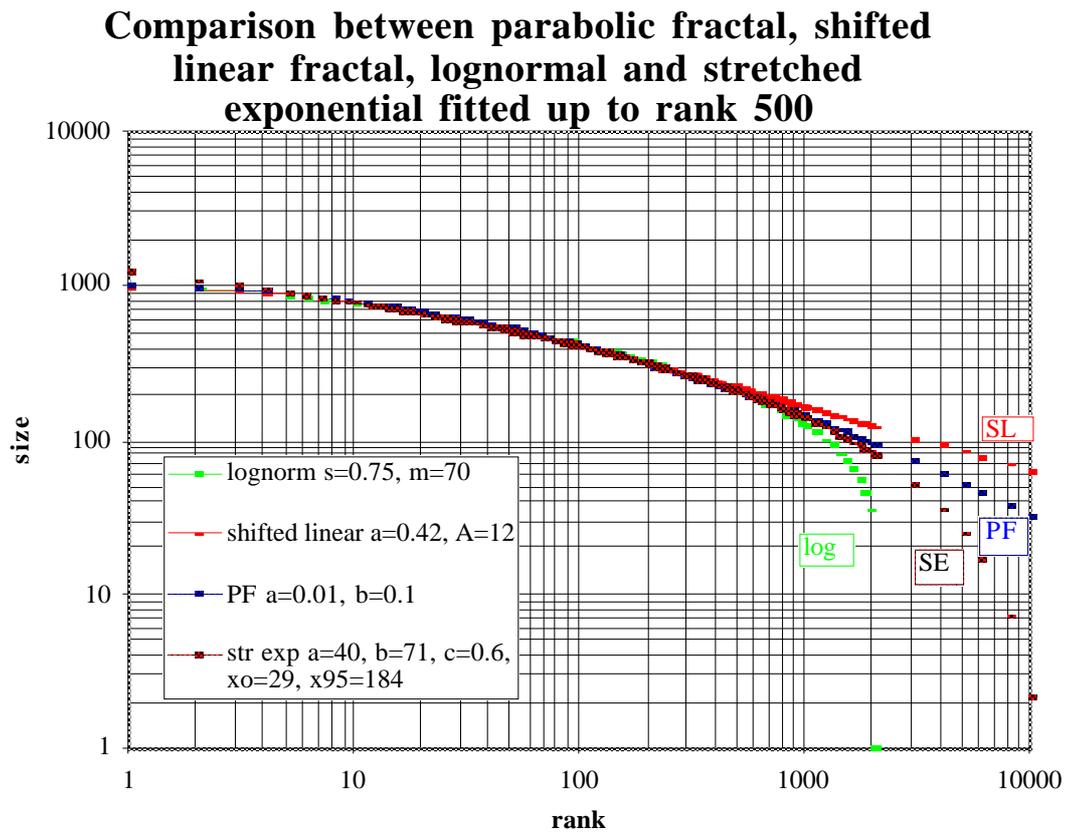

**figure 2**

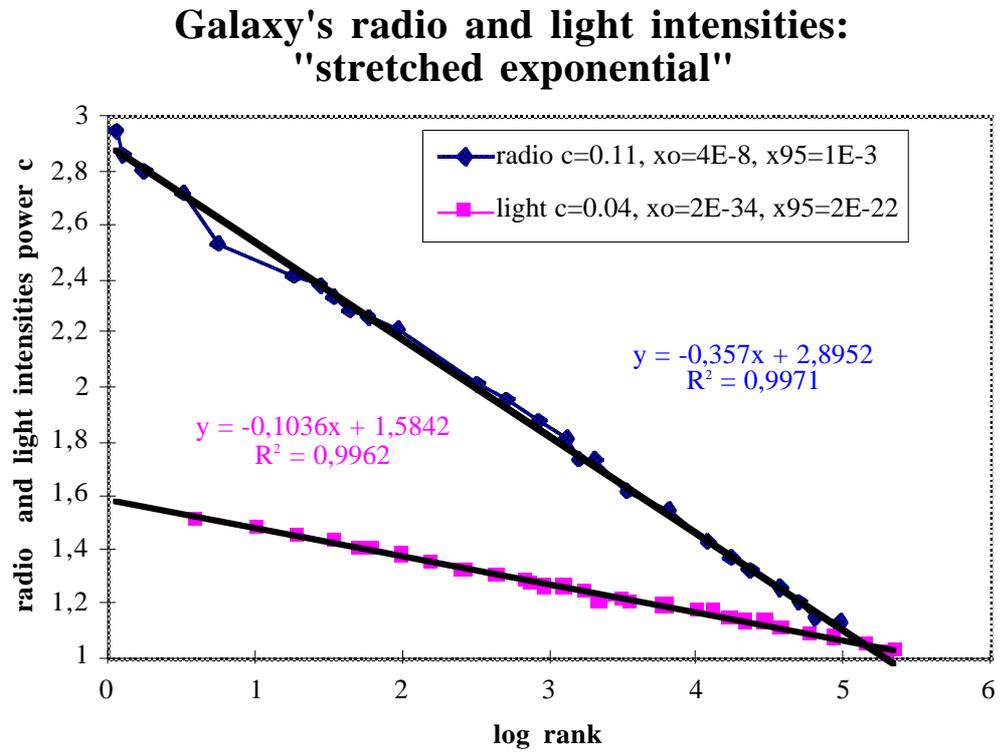

**Galaxy's radio and light intensities:
"stretched exponential"**

**figure 3 :**

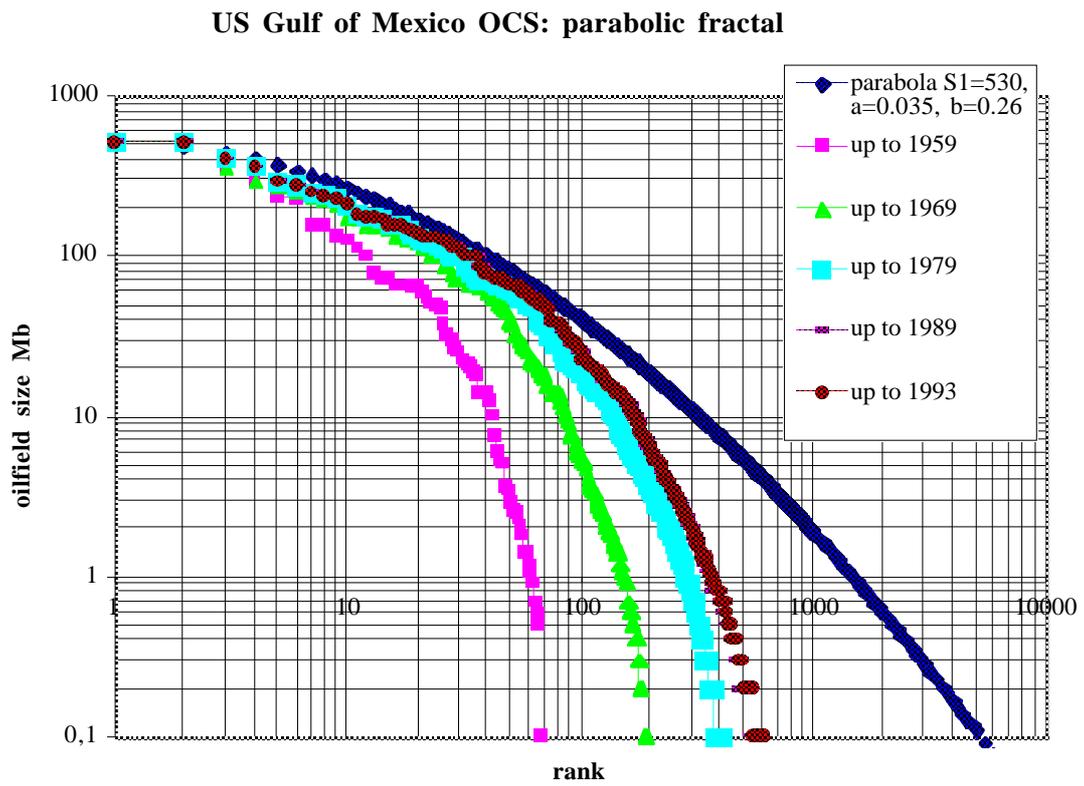

US Gulf of Mexico OCS: parabolic fractal

**figure 4 :**

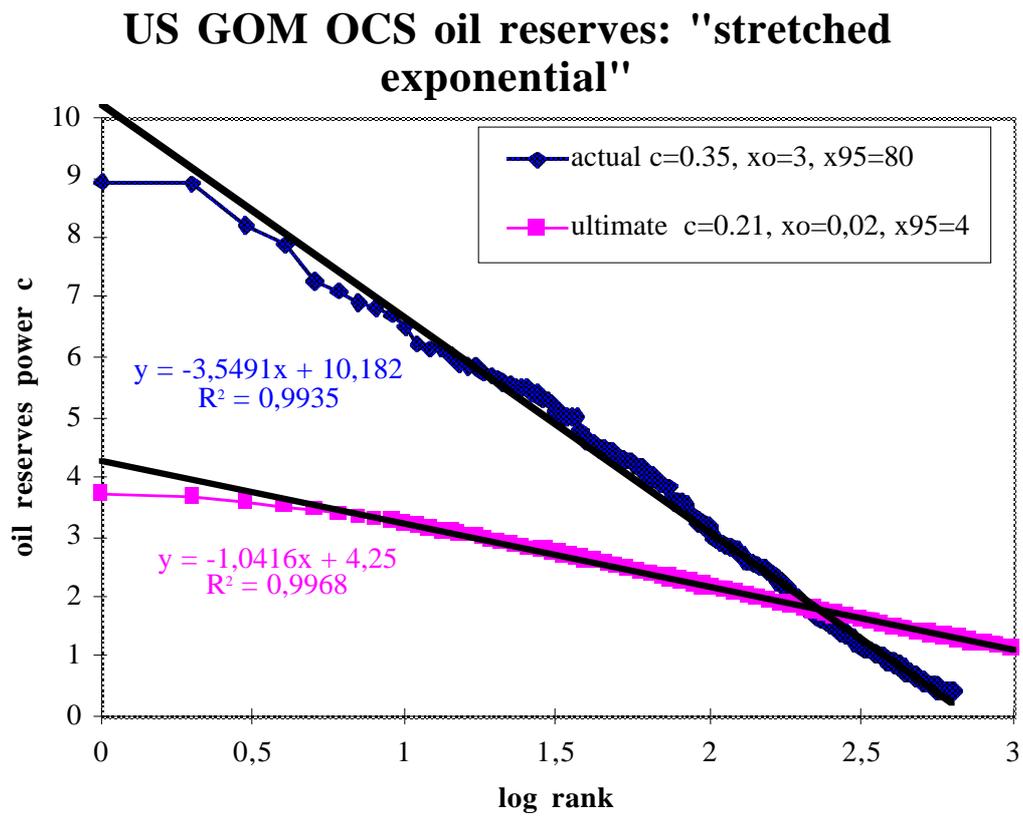

US GOM OCS oil reserves: "stretched exponential"

**figure 5 :**

### US urban agglomerations >100 000 inhabitants: "stretched exponential"

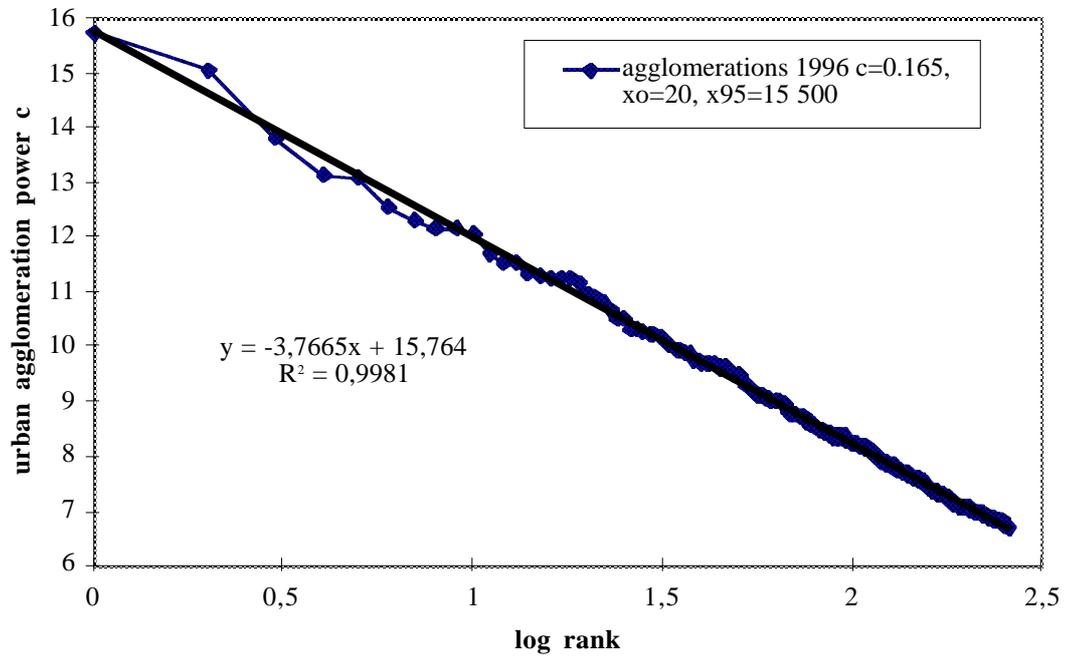

**figure 6 :**

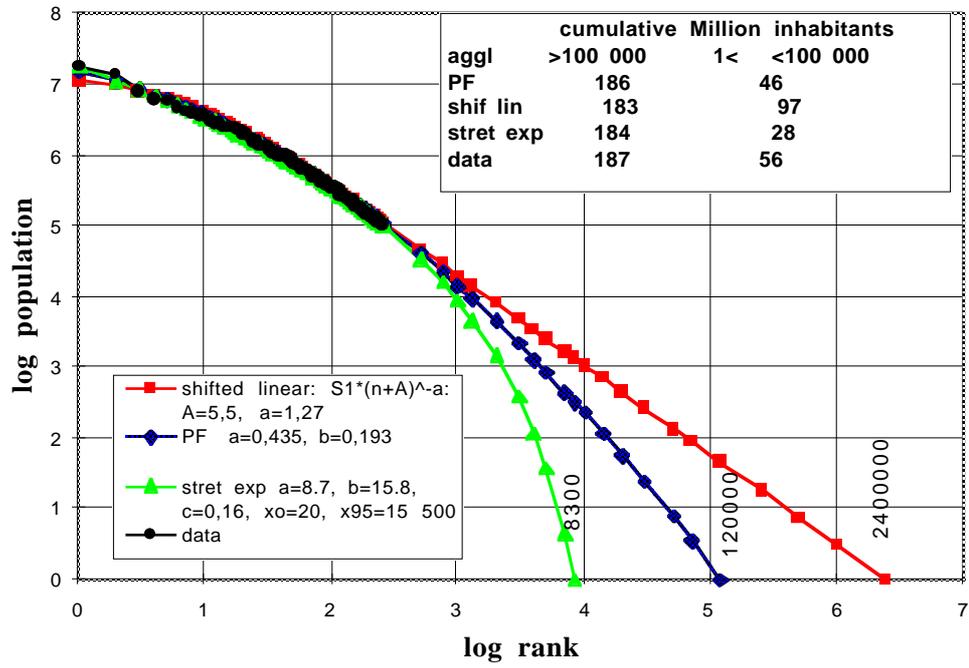

## US agglomerations: parabolic, shifted linear fractal and stretched exponential

**figure 7 :**

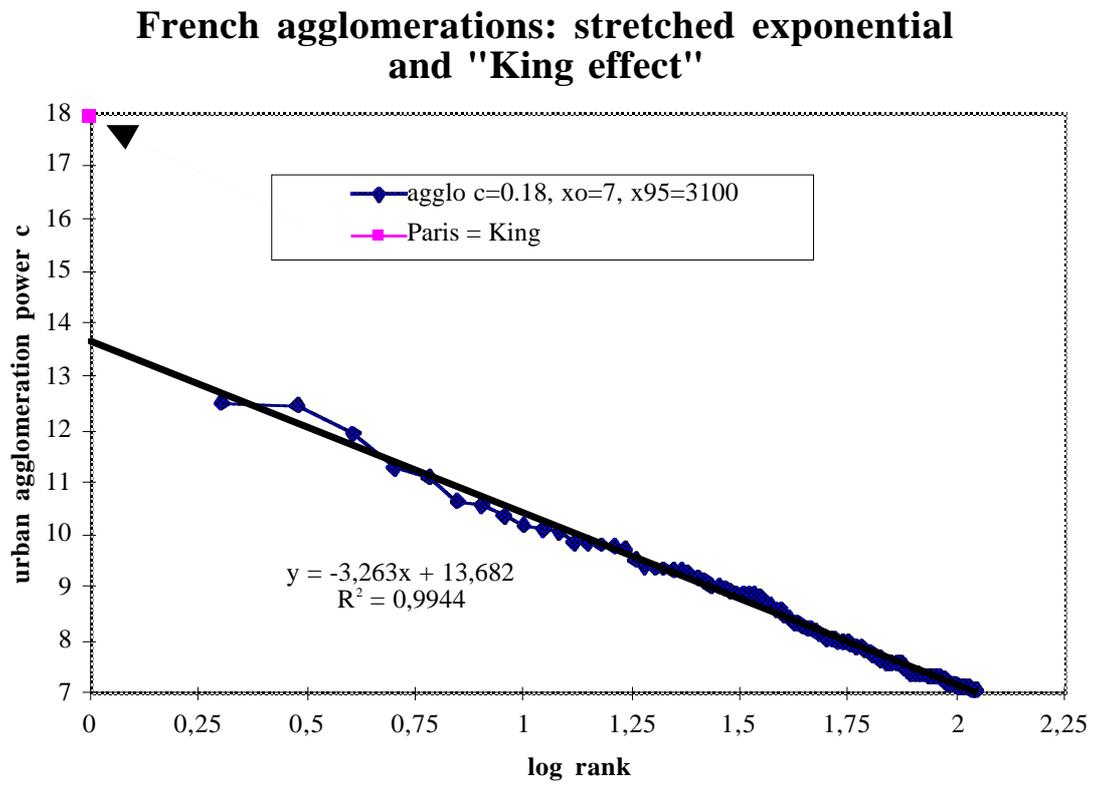

**figure 8 :**

## World's urban agglomeration: "stretched exponential"

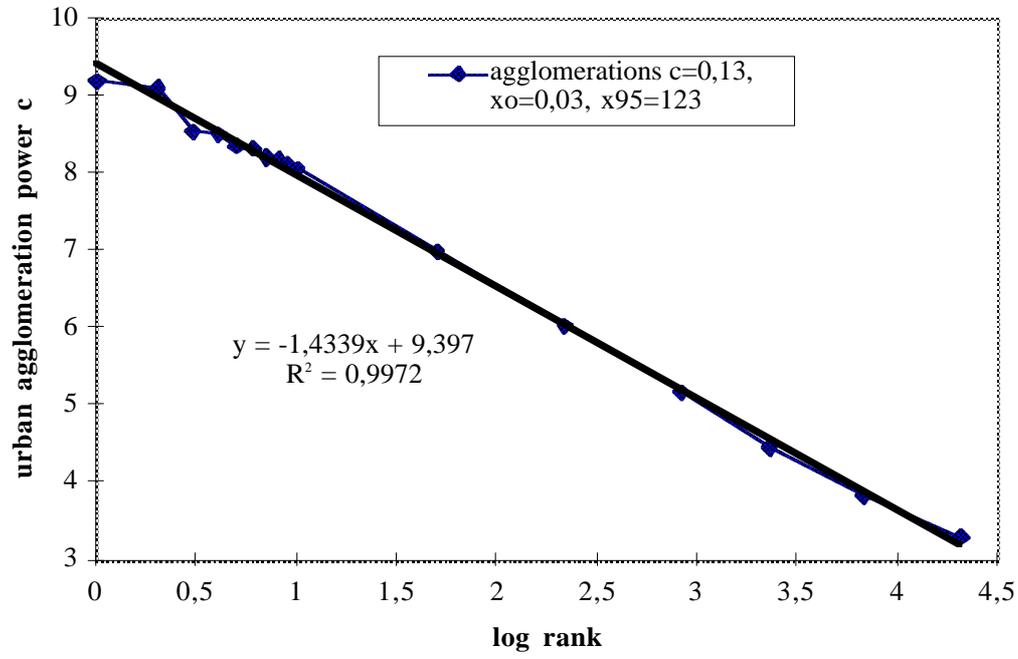

**Figure 9 :**

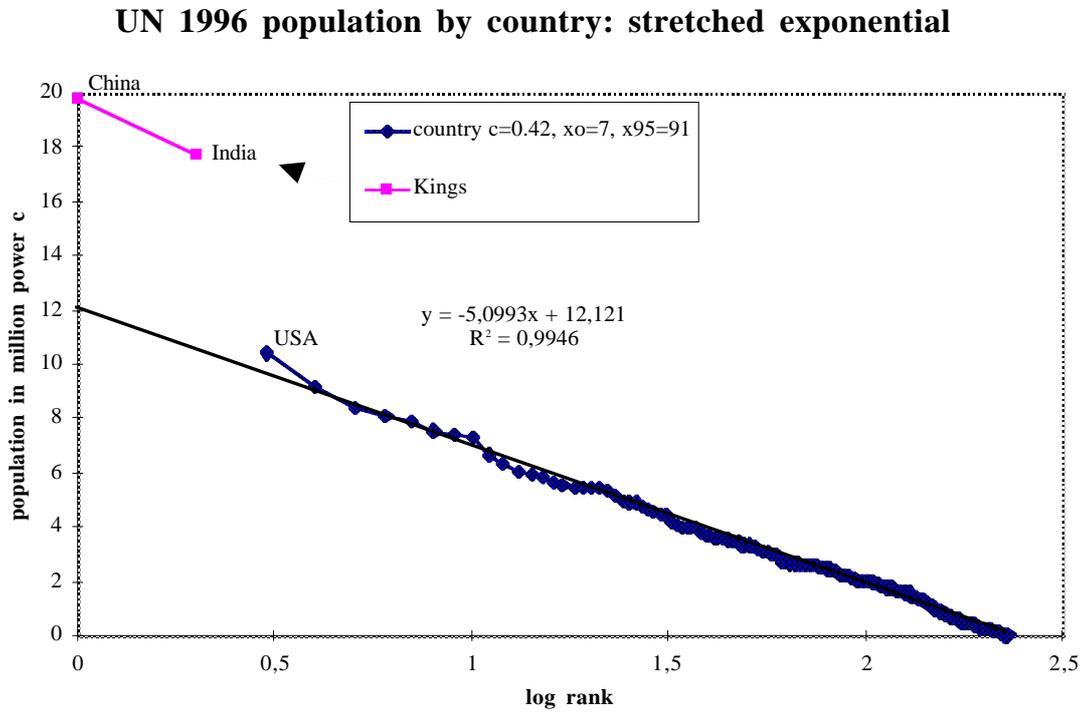

**UN 1996 population by country: stretched exponential**

**Figure 10a :**

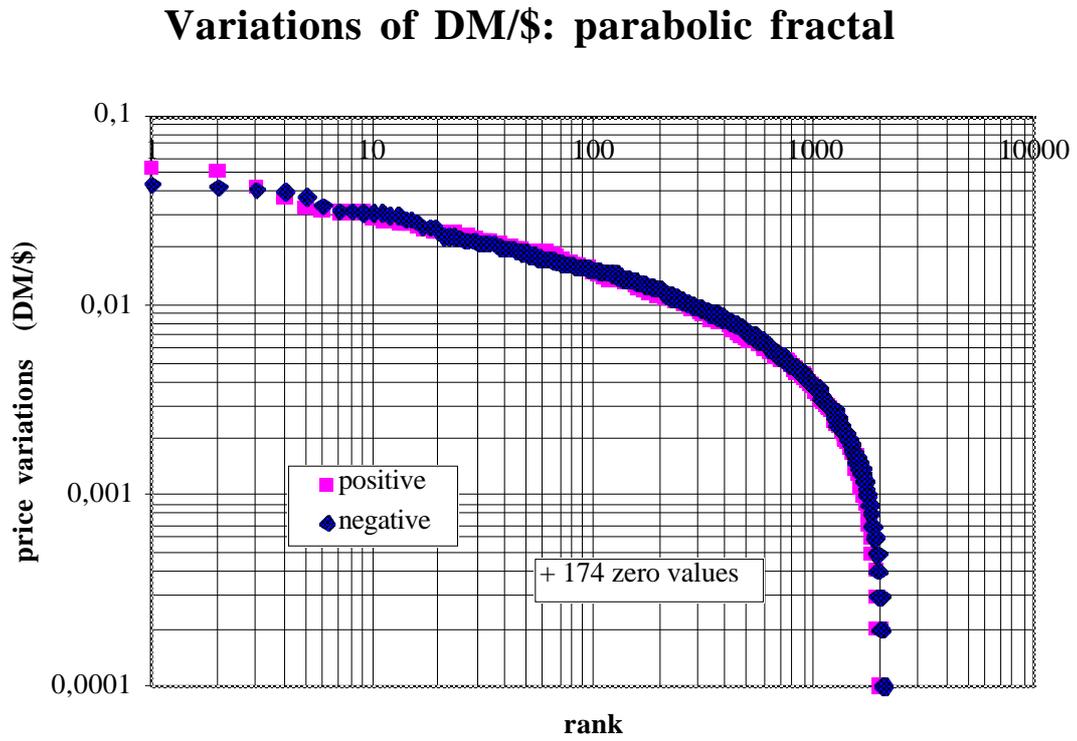

**Variations of DM/$: parabolic fractal**

**Figure 10b :**

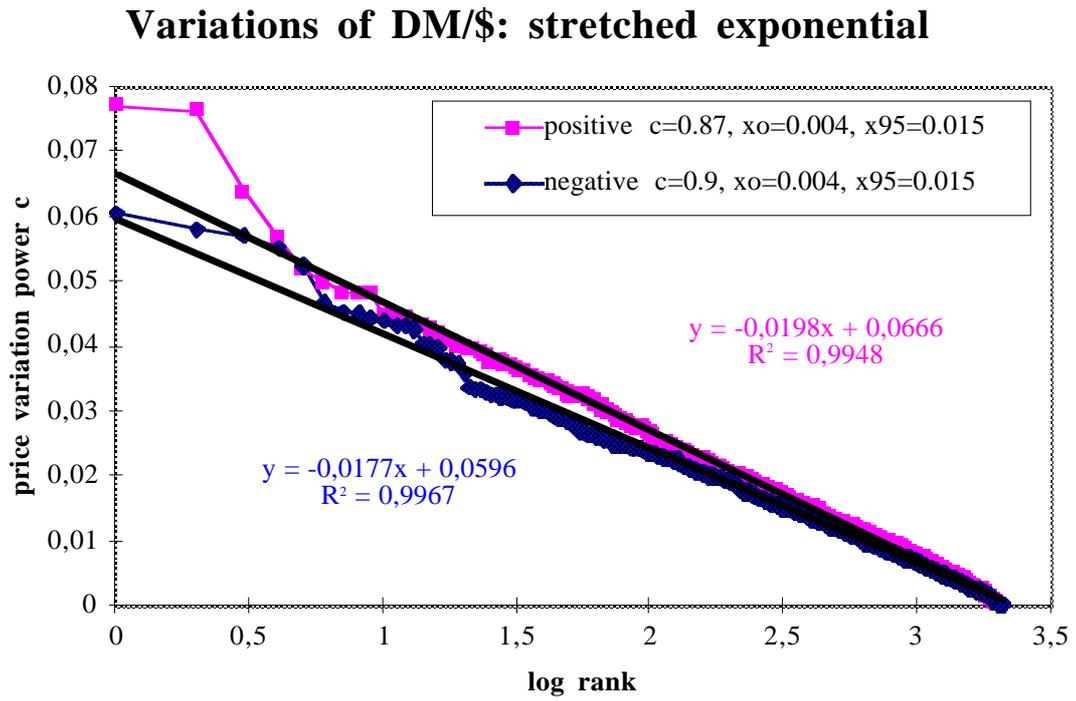

**Variations of DM/$: stretched exponential**

Legend:
- positive  c=0.87, xo=0.004, x95=0.015
- negative  c=0.9, xo=0.004, x95=0.015

y = -0,0198x + 0,0666
R² = 0,9948

y = -0,0177x + 0,0596
R² = 0,9967

x-axis: **log rank**
y-axis: **price variation power c**

**Figure 11a :**

## variations of M/F: Parabolic fractal

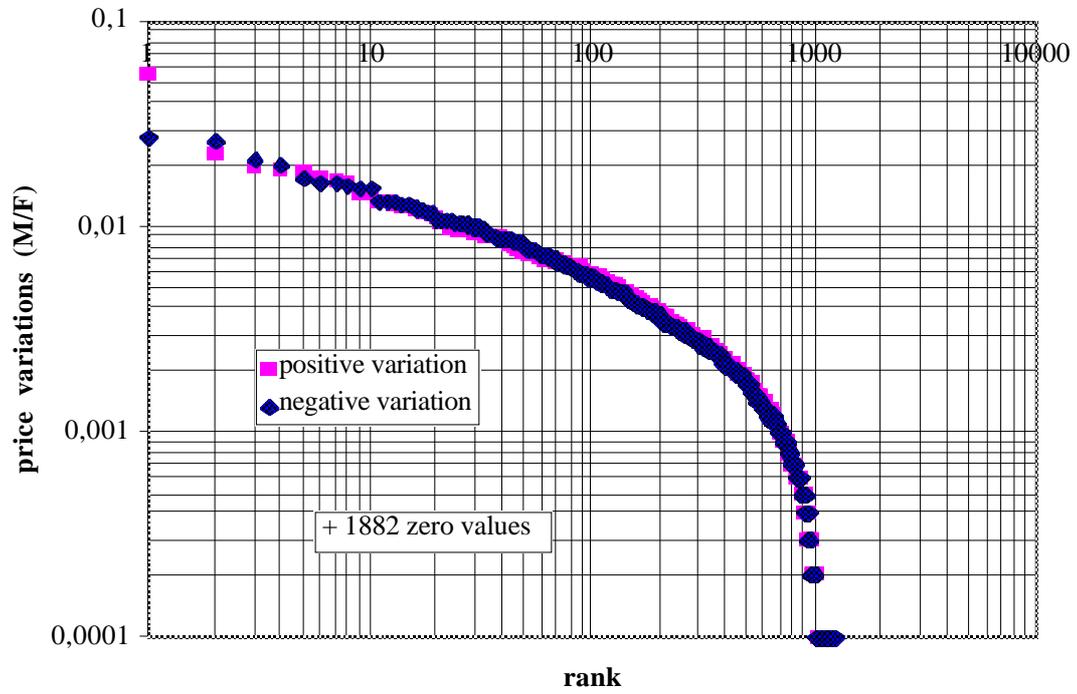

**Figure 11b :**

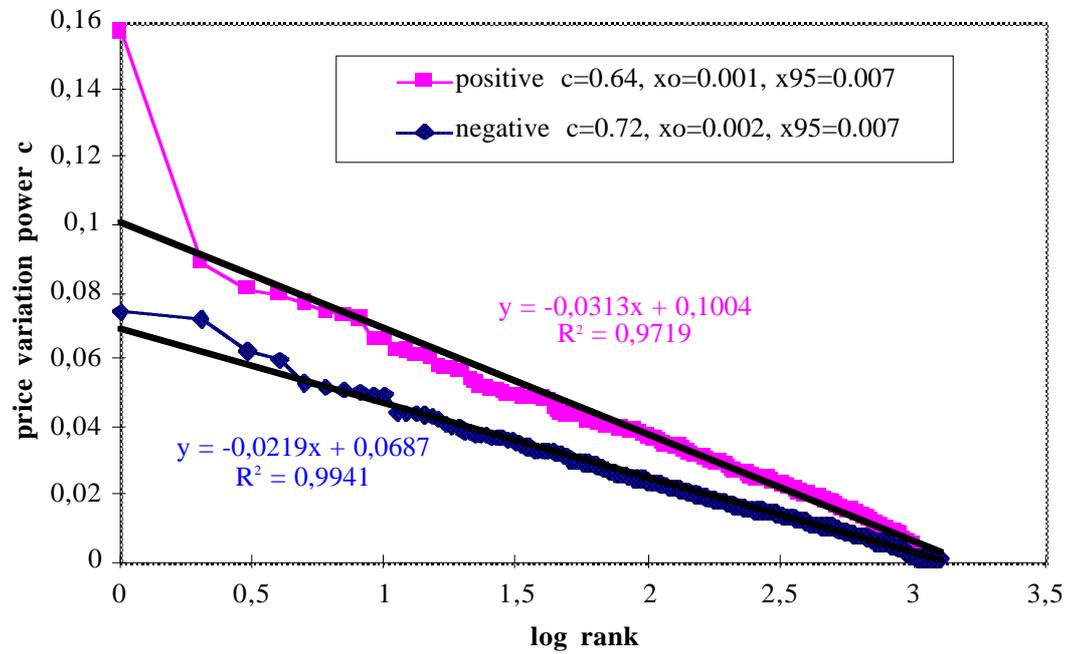

## Variations of Mark/Franc: stretched exponential

**Figure 12a :**

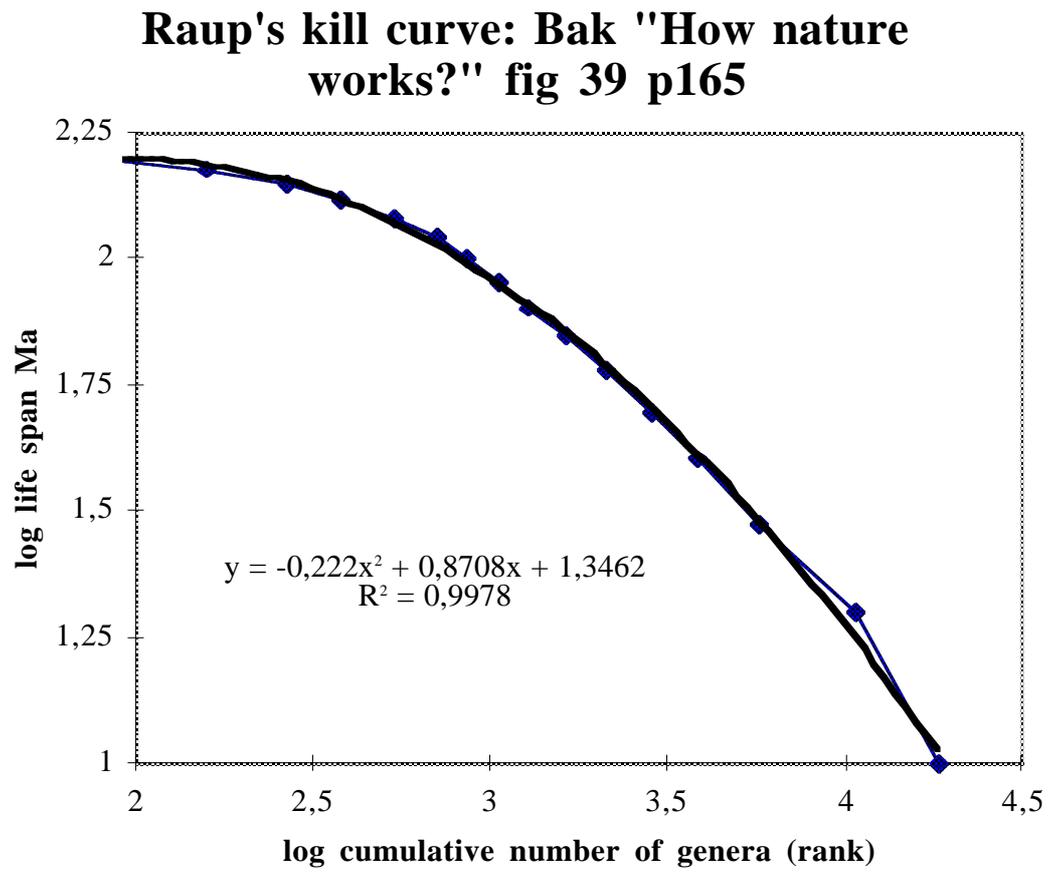

Raup's kill curve: Bak "How nature works?" fig 39 p165

$y = -0,222x^2 + 0,8708x + 1,3462$
$R^2 = 0,9978$

log life span Ma

log cumulative number of genera (rank)

**figure 12b :**

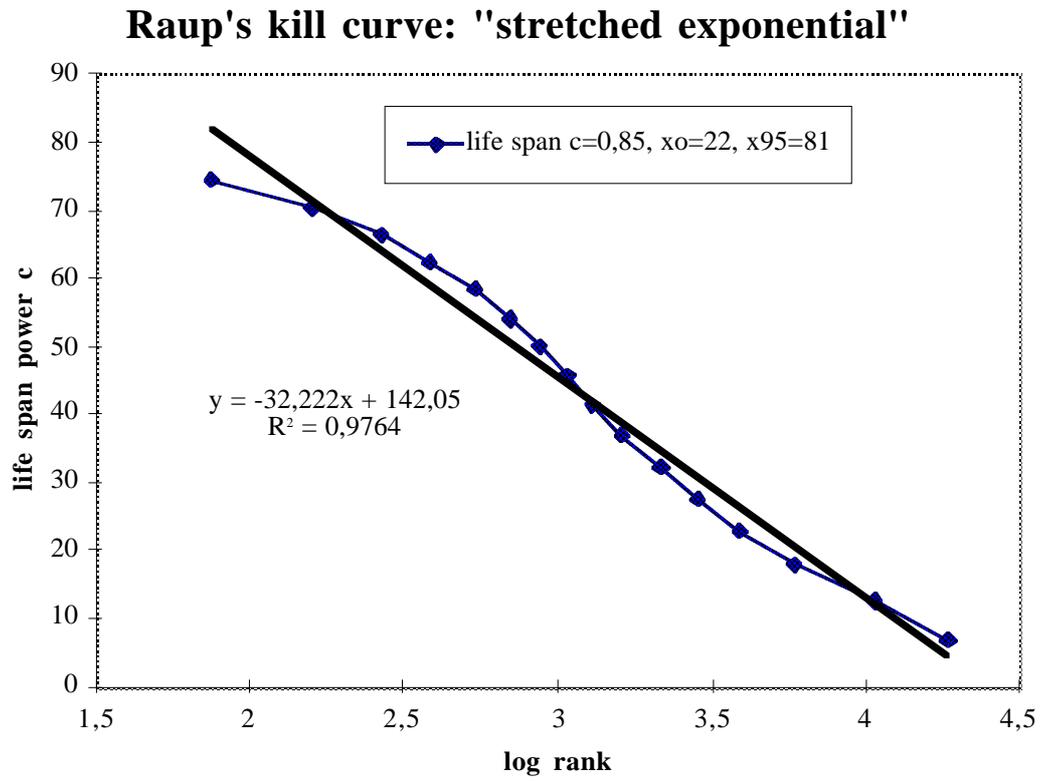

**Raup's kill curve: "stretched exponential"**

$y = -32,222x + 142,05$
$R^2 = 0,9764$

life span c=0,85, xo=22, x95=81

life span power c

log rank

**figure 13a :**

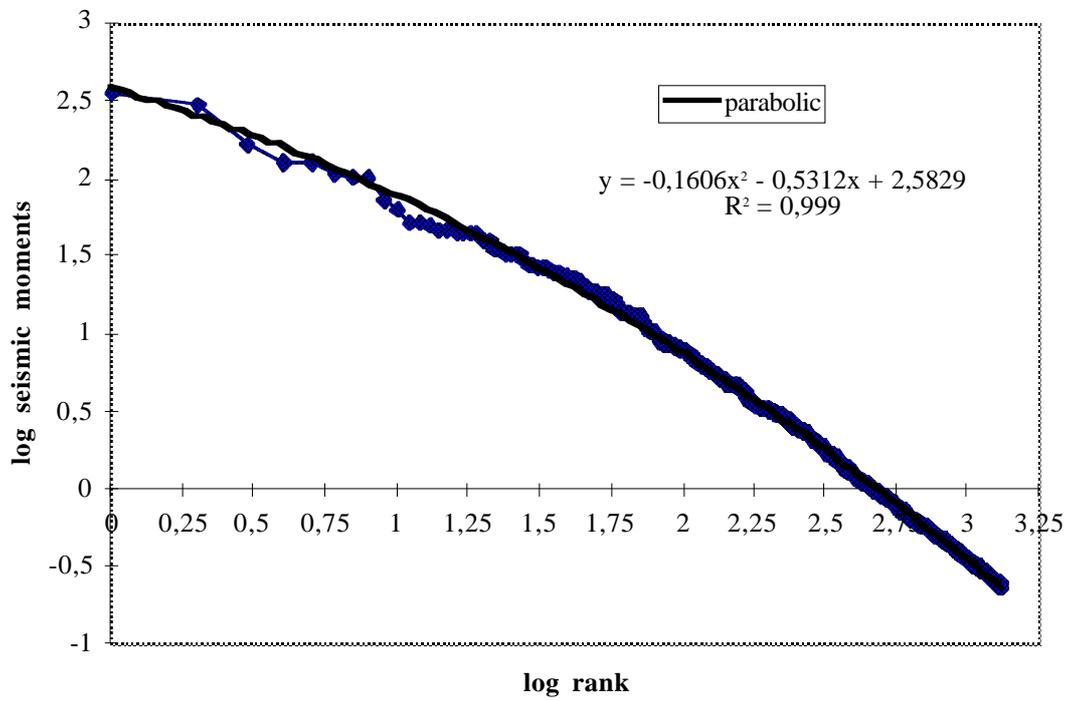

**seismic moments**

$y = -0,1606x^2 - 0,5312x + 2,5829$
$R^2 = 0,999$

**figure 13b :**

## seismic moments: stretched exponential

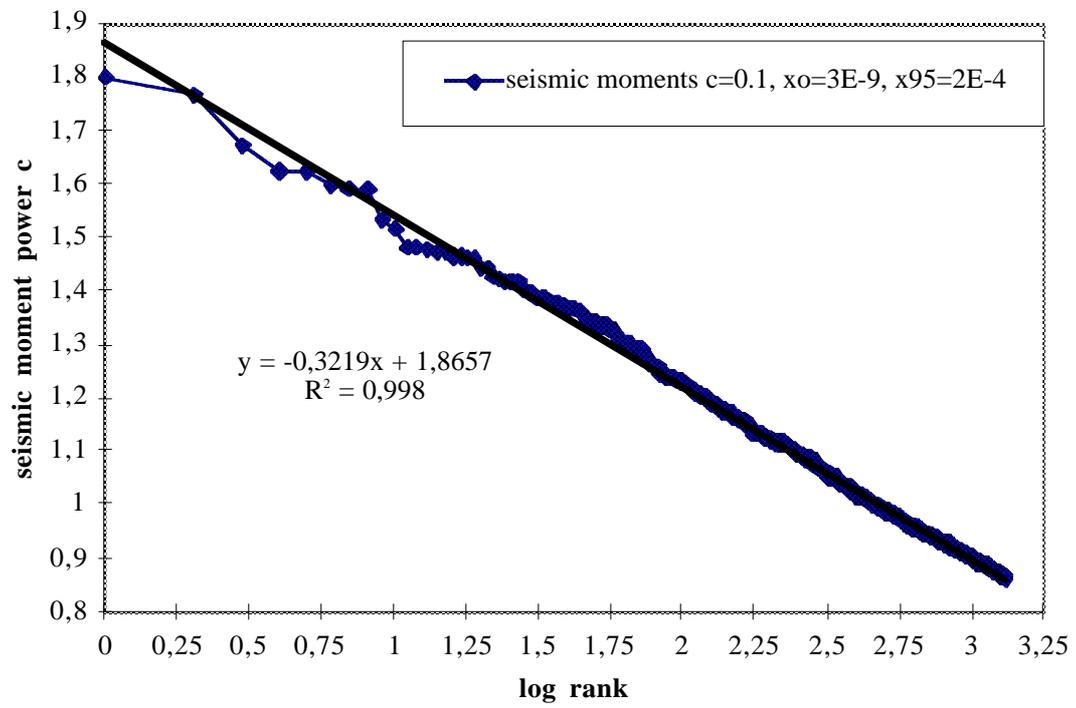

**Figure 14a :**

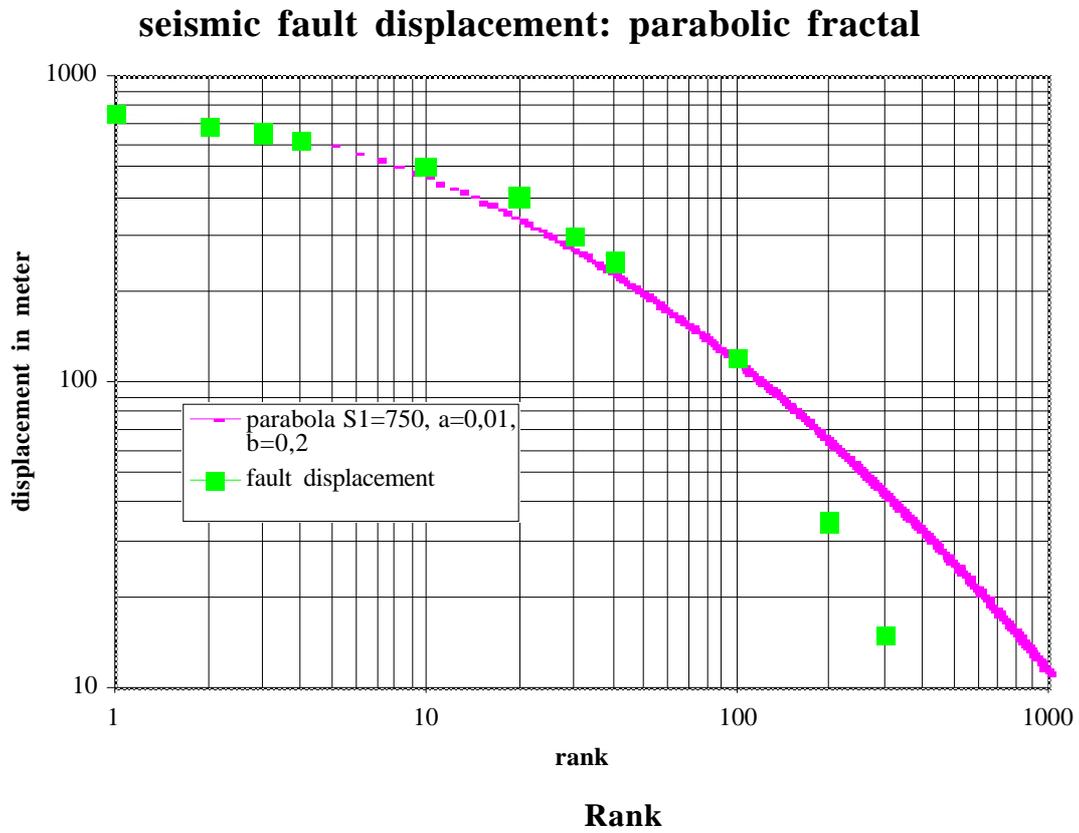

seismic fault displacement: parabolic fractal

**Rank**

**Figure 14b :**

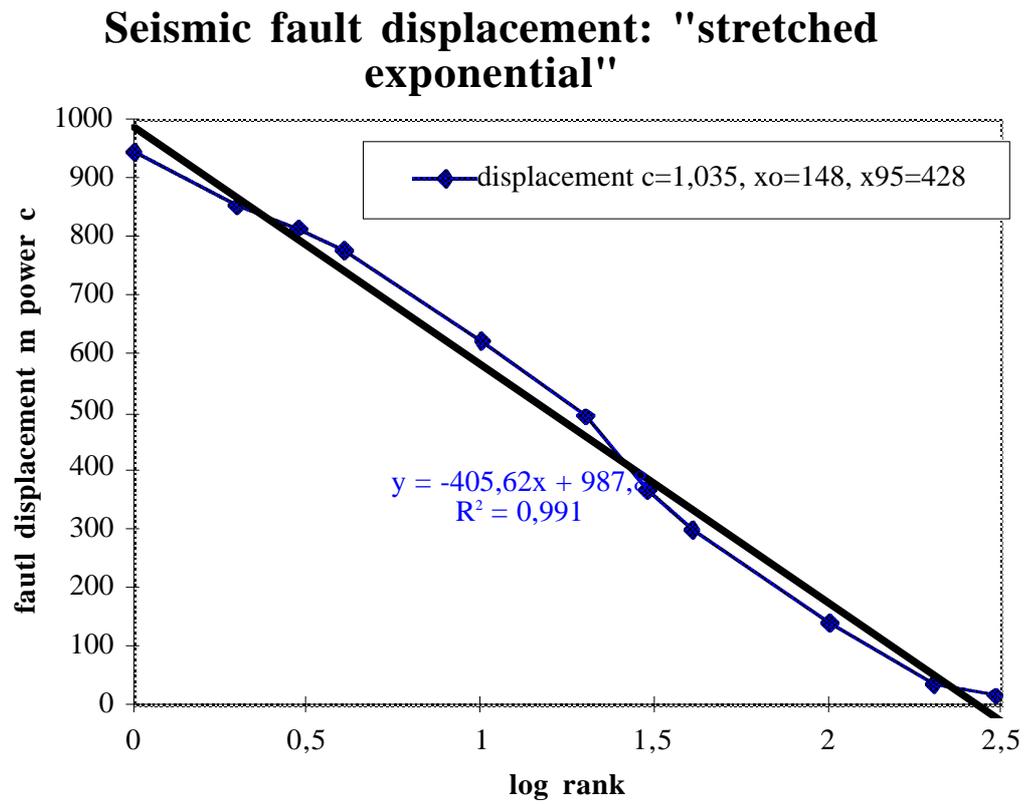

**Seismic fault displacement: "stretched exponential"**

displacement c=1,035, xo=148, x95=428

y = -405,62x + 987,8
R² = 0,991

fautl displacement m power c

log rank

**Figure 15a**

### Vostok: variations of temperatures versus time: parabolic fractal

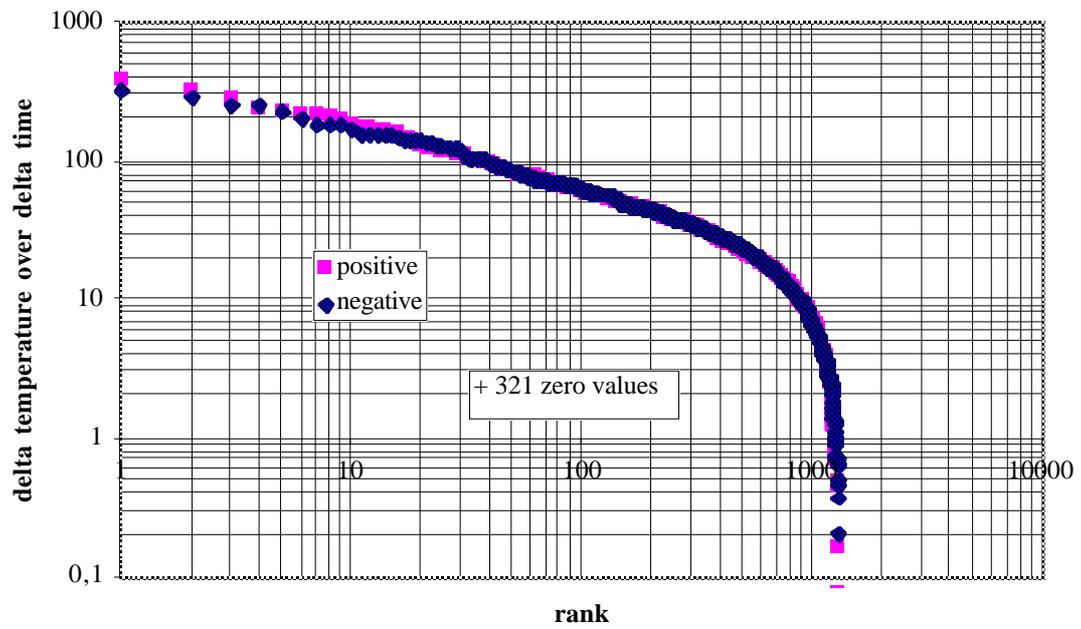

**Figure 15b**

## Vostok: variations temperatures vs time; stretched exponential

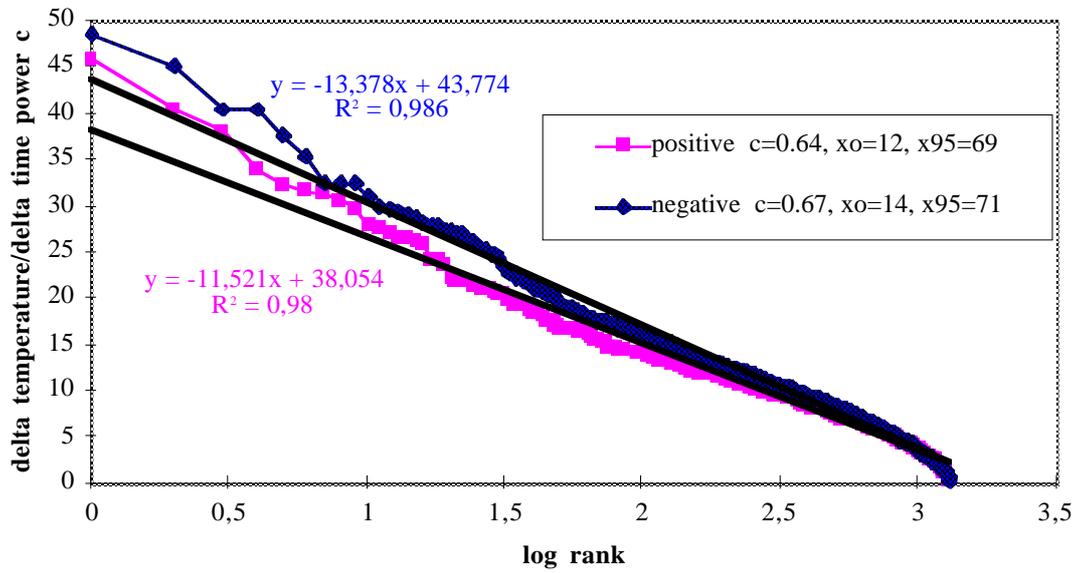

**Figure 16a**

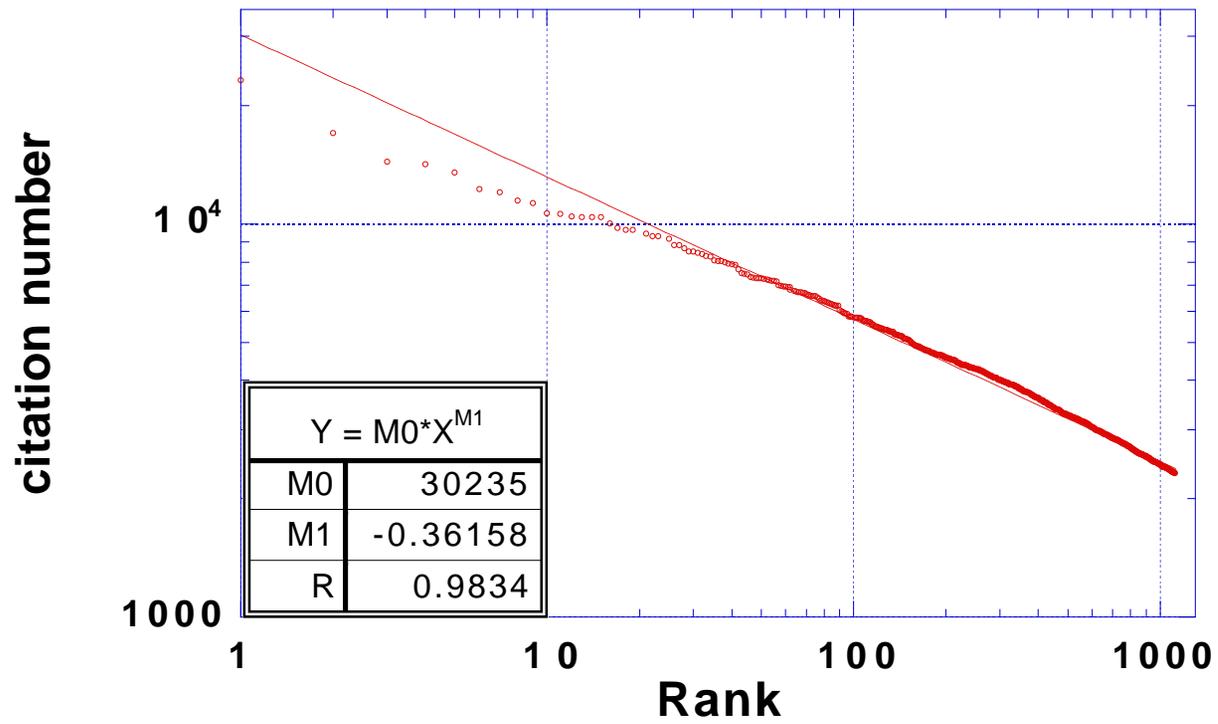

**Figure 16b**

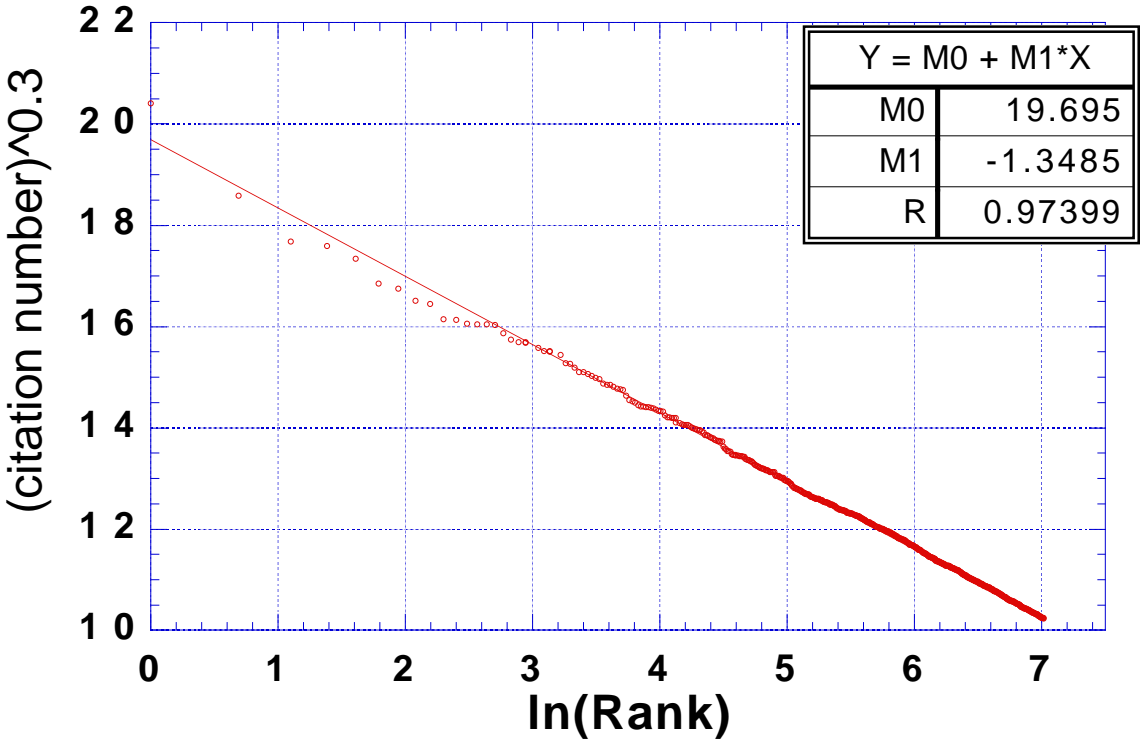